\documentclass[aps,prl,showpacs,twocolumn]{revtex4}
\usepackage{amsmath}
\usepackage{graphicx}
\usepackage{dsfont}

\begin{document}

\title{Quantum antiferromagnetic Heisenberg half-odd integer spin model as
the entanglement Hamiltonian of the integer spin Affleck-Kennedy-Lieb-Tasaki
states}
\author{Wen-Jia Rao$^{1}$ and Guang-Ming Zhang$^{1,2}$}
\email{gmzhang@tsinghua.edu.cn}
\author{Kun Yang$^{3}$}
\affiliation{$^{1}$State Key Laboratory of Low-Dimensional Quantum Physics and Department
of Physics, Tsinghua University, Beijing 100084, China. \\
$^{2}$Collaborative Innovation Center of Quantum Matter, Beijing 100084,
China.\\
$^{3}$National High Magnetic Field Laboratory and Physics Department,
Florida State University, Tallahassee, Florida 32310, USA.}
\date{\today}

\begin{abstract}
Applying a symmetric bulk bipartition to the one-dimensional
Affleck-Kennedy-Lieb-Tasaki valence bond solid (VBS) states for the integer
spin-S Haldane gapped phase, we can create an array of fractionalized
spin-S/2 edge states with the super unit cell $l$ in the reduced bulk
system, and the topological properties encoded in the VBS wave functions can
be revealed. The entanglement Hamiltonian (EH) with $l=even$ corresponds to
the quantum antiferromagnetic Heisenberg spin-S/2 model. For the even
integer spins, the EH still describes the Haldane gapped phase. For the odd
integer spins, however, the EH just corresponds to the quantum
antiferromagnetic Heisenberg half-odd integer spin model with spinon
excitations, characterizing the critical point separating the topological
Haldane phase from the trivial gapped phase. Our results thus demonstrate
that the topological bulk property not only determines its fractionalized
edge states, but also the quantum criticality associated with the
topological phase, where the elementary excitations are precisely those
fractionalized edge degrees of freedom \emph{confined} in the bulk of the
topological phase.
\end{abstract}

\maketitle

\section{Introduction}

Topological phases of matter including those require symmetry protection
have been the subject of intense interest in quantum information science,
condensed matter physics and quantum field theory. Much effort has been
devoted to classification of these topological phases, and tremendous
success is achieved in our understanding of quantum Hall states\cite{TKNN},
topological insulators\cite{Hasan-Kane,Qi-Zhang,Moore}, and symmetry
protected topological (SPT) phases\cite%
{Chen-Gu-Wen,Schuch,Chen-Gu-Liu-Wen-2013}. The SPT phases possess bulk
energy gaps and do not break any symmetry, but have robust gapless edge
excitations. These SPT states can not be continuously connected to a trivial
gapped state without closing the energy gap. So there exists a topological
phase transition between a SPT phase and its adjacent trivial phase, and the
corresponding critical theory does not belong to the conventional
Landau-Ginzburg-Wilson paradigm\cite%
{Chen-Wang-Lu-Lee,Rao-Wan-Zhang,Hsieh-Fu-Qi,Tsui-Lu-Jiang-Lee}. Such a
critical point is a prototype of ``deconfined quantum critical point (QCP)''
with fractionalized elementary excitations\cite{Senthil}. A crucial question
is how to extract the critical properties from the ground state wave
function of the SPT phases.

In one dimension, Haldane\cite{Haldane-1983} predicted that quantum
antiferromagnetic Heisenberg spin chains are classified into two
universality classes: half-odd integer spins with gapless excitations and
integer spins with gapped excitations. Recent studies\cite%
{Gu-Wen-2009,Pollmann-2012} indicated that the Haldane gapped phase for
\textit{odd} integer spin chains is a typical SPT phase, while the \textit{%
even} integer spin chains correspond to the topologically trivial phase,
because their edge states are not protected by the projective representation
of the bulk SO(3) symmetry. According to the classification theory\cite%
{Chen-Gu-Liu-Wen-2013}, there exists only one nontrivial SPT phase for the
SO(3) symmetric quantum Heisenberg spin model, whose fixed point wave
function is given by the Affleck-Kennedy-Lieb-Tasaki (AKLT) valence bond
solid state (VBS)\cite{AKLT}. Since the symmetry protection of the SPT phase
in the bulk can be analyzed in terms of symmetry protection of the
fractionalized edge spins, it motivates us to question if there exists a
general connection between the SPT phase and the quantum critical phases of
the quantum antiferromagnetic Heisenberg half-odd-integer spin chains.

In this paper, we first review the entanglement property of a single block
in the one-dimensional integer spin-S AKLT VBS states, and prove that the
entanglement Hamiltonian can be expressed in terms of the Heisenberg
exchange of two edge spin-S/2's. By using a symmetric bulk bipartition\cite%
{Rao-Wan-Zhang,rao-cai-wan-zhang,santos-lundgren}, we can create an array of
fractionalized spin-S/2 edge states with super unit cell $l$ in the reduced
bulk system. Then the reduced density matrix and entanglement Hamiltonian
(EH) can be derived in terms of the fractionalized edge spins, leading to
the quantum antiferromagnetic Heisenberg spin-S/2 model when the super unit
cell $l$ includes even number of lattice sites. For $S=4n+2$ with integer $n$%
, the EH still describes the nontrivial Haldane gapped phase with odd
integer spins, and for $S=4n$ the EH corresponds to the even integer Haldane
gapped phase. For the odd integer spin-$S$, however, the quantum
antiferromagnetic Heisenberg half-odd integer spin model emerges,
characterizing the quantum critical point separating the nontrivial Haldane
phase from the trivial phase. So our results demonstrate that the
topological bulk property not only determines its fractionalized edge
states, but also the critical point at the continuous phase transition to
its nearby trivial phase.

\section{Single block entanglement}

\begin{figure}[t]
\includegraphics[width=8cm]{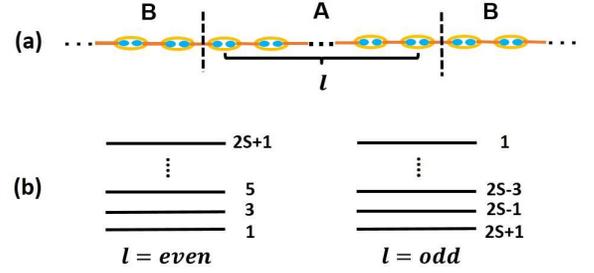}
\caption{(a) The picture of AKLT VBS state. Each blue dot represents a spin-$%
S/2$, yellow circle stands for the physical spin-$S$, and solid lines denote
the singlet bonds. A block with $l$ sites is chosen as the subsystem A. (b)
The entanglement spectra of the single block are given for $l=$even and $l=$%
odd, respectively.}
\label{fig:SingleBlockES}
\end{figure}

The spin-$S$ AKLT VBS state as the fixed point state of the Haldane gapped
phase is defined by%
\begin{equation}
|\text{VBS}\rangle =\prod_{i=0}^{N}\left( a_{i}^{\dagger }b_{i+1}^{\dagger
}-b_{i}^{\dagger }a_{i+1}^{\dagger }\right) ^{S}|vac\rangle ,
\end{equation}%
where $a_{i}^{\dagger }$ and $b_{i}^{\dagger }$ are the Schwinger boson
creation operators with a local constraint $a_{i}^{\dagger
}a_{i}+b_{i}^{\dagger }b_{i}=2S$, and the spin operators are expressed as $%
S_{i}^{+}=a_{i}^{\dagger }b_{i}$, $S_{i}^{-}=b_{i}^{\dagger }a_{i}$, and $%
S_{i}^{z}=\left( a_{i}^{\dagger }a_{i}-b_{i}^{\dagger }b_{i}\right) /2$. In
this construction, each physical spin is composed of two spin-$S/2$'s
projected into a total spin-$S$ state, while each neighboring sites are
linked by spin-$S/2$ singlet, see Fig.~\ref{fig:SingleBlockES}(a).\

To consider the entanglement properties, we choose a block of $l$ sites
denoted by the part A. With the help of the spin coherent state
representation, the reduced density matrix $\rho _{A}$ can be obtained by
tracing out the degrees of freedom without the part A, and its nonzero
eigenvalues $\lambda _{j}$ with degeneracy $2j+1$ have been derived\cite%
{Hosho,Xu}
\begin{eqnarray}
\lambda _{j} &=&\frac{1}{\left( S+1\right) ^{2}}\sum_{k=0}^{S}\left(
2k+1\right) \left[ f(k)\right] ^{l-1}  \notag \\
&&\times I_{k}\left[ \frac{1}{2}j\left( j+1\right) -\frac{1}{4}S\left(
S+2\right) \right] , \\
f(k) &=&\frac{\left( -1\right) ^{k}S!\left( S+1\right) !}{\left( S-k\right)
!\left( S+k+1\right) !},
\end{eqnarray}%
where $j=0,1,...S$ and the recursion function $I_{k}[x]$ is defined by
\begin{eqnarray}
I_{k+1}[x] &=&\frac{2k+1}{\left( S+k+2\right) ^{2}}\left( k+\frac{4x}{k+1}%
\right) I_{k}[x]  \notag \\
&&-\frac{k}{k+1}\left( \frac{S-k+1}{S+k+2}\right) ^{2}I_{k-1}[x],
\end{eqnarray}%
with $I_{0}[x]=1$ and $I_{1}[x]=\frac{4x}{\left( S+2\right) ^{2}}$. Since
the function $\left| f\left( k\right) \right| $ decreases with $k$ very
quickly, only the first two terms ($k=0,1$) dominate in the summation for a
long block length $l$. Thus the eigenvalues are approximated as
\begin{equation*}
\lambda _{j}\approx \frac{1}{\left( S+1\right) ^{2}}+3\left( \frac{-S}{S+2}%
\right) ^{l-1}\frac{\left[ 2j\left( j+1\right) -S\left( S+2\right) \right] }{%
\left( S+2\right) ^{4}},
\end{equation*}%
and up to the first order of $\delta =\left( \frac{-S}{S+2}\right) ^{l}$ the
entanglement spectrum is thus derived as{\small
\begin{equation}
\xi _{j}\approx J\left( l\right) \left[ \frac{1}{2}j\left( j+1\right) -\frac{%
S}{2}\left( \frac{S}{2}+1\right) \right] ,
\end{equation}%
} with $J\left( l\right) =\frac{12}{S\left( S+2\right) }\left( \frac{-S}{S+2}%
\right) ^{l}$. Then the corresponding EH can be recognized as: $%
H_{E}=J\left( l\right) \mathbf{s}_{1}\cdot \mathbf{s}_{2}$ where $\mathbf{s}%
_{1}$ and $\mathbf{s}_{2}$ are the fractionalized edge spins. Therefore, for
a long block length $l$, the entanglement properties of the single block are
just described by the quantum Heisenberg spin model, and the corresponding
entanglement spectra are displayed in Fig.~\ref{fig:SingleBlockES}(b).

\section{Symmetric bulk bipartition}

The symmetric bulk bipartition is the most effective tool to generate an
extensive array of fractionalized edge spin-S/2's in the bulk subsystem,
i.e., the spin chain is divided into two subsystems both including the same
number of disjoint blocks\cite{Rao-Wan-Zhang,rao-cai-wan-zhang}. The
fractionalized edge spins can thus percolate in the reduced bulk system and
emerge as coherent elementary excitations of the effective field theory of
the subsystem. It is convenient to write the AKLT VBS wave function in the
form of matrix product state (MPS) representation shown in Fig.~\ref%
{fig:MPS-Graph}(a)%
\begin{equation}
|VBS\rangle =\sum_{\left\{ s_{i}\right\} }\text{Tr}\left[ A^{\left[ s_{1}%
\right] }A^{\left[ s_{2}\right] }..A^{\left[ s_{N}\right] }\right]
|s_{1},s_{2},..s_{N}\rangle ,
\end{equation}%
where $A^{\left[ s_{i}\right] }$ are the $\left( S+1\right) \times \left(
S+1\right) $ local matrices, whose elements can be obtained from the
Schwinger boson representation, and the periodic boundary condition are
assumed. When we group each continuous $l$ lattice sites into a block, all
the even blocks are denoted by the part A and the rest by the part B. Then
by tracing out the part B, the reduced density matrix $\rho _{A}$ and the EH
($H_{E}=-\ln \rho _{A}$) can be derived. The general procedure is described
as the following four steps.

\begin{figure}[t]
\includegraphics[width=8cm]{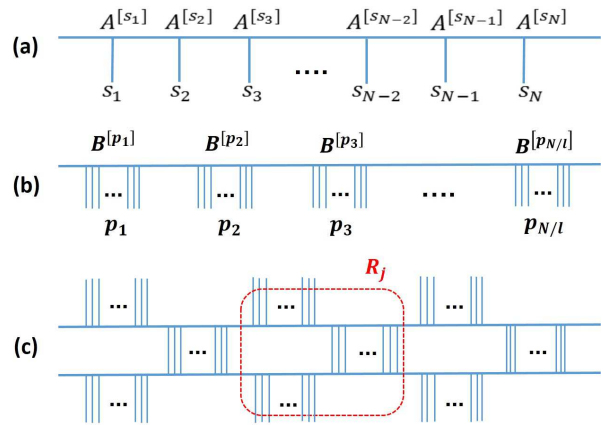} 
\caption{(a) The MPS representation of the AKLT VBS state. (b) The blocked
AKLT VBS state. (c) The reduced density matrix under symmetric bulk
bipartition with a repeating structure.}
\label{fig:MPS-Graph}
\end{figure}

Step 1. Conduct the coarse graining and distill relevant states within each
block\cite{Cirac}. We pick out a block with $l$ sites, and perform the
singular value decomposition%
\begin{equation}
\left( A^{\left[ s_{1}\right] }A^{\left[ s_{2}\right] }..A^{\left[ s_{l}%
\right] }\right) _{\alpha ,\beta }=\sum_{p=0}^{\kappa -1}X_{\left( \left\{
s_{i}\right\} \right) ,p}\Lambda _{p}Y_{p,\left( \alpha ,\beta \right) },
\end{equation}%
where the number of nonzero singular values $\kappa $ records the number of
relevant states in the block. For the spin-$S$ AKLT VBS state, $\kappa
=\left( S+1\right) ^{2}$, and the relevant states $|p\rangle $ are
effectively composed by two edge spin-$S/2$'s:
\begin{equation*}
|p\rangle =\sum_{m,n}\chi _{m,n}^{p}|m,n\rangle ,
\end{equation*}%
which are the combination of the degenerate edge states $|m,n\rangle $ with $%
m,n\in \left[ -S/2,S/2\right] $. Then we can rewrite the original VBS wave
function into the blocked MPS form, see Fig.~\ref{fig:MPS-Graph}(b)%
\begin{equation}
|\Psi \rangle =\sum_{\left\{ p_{i}\right\} }\text{Tr}\left( B^{\left[ p_{1}%
\right] }B^{\left[ p_{2}\right] }..B^{\left[ p_{N/l}\right] }\right)
|p_{1},p_{2},..p_{N/l}\rangle ,
\end{equation}%
where the block matrices are given by $B_{\alpha ,\beta }^{\left[ p\right]
}=\Lambda _{p,p}Y_{p,\left( \alpha ,\beta \right) }$.

Step 2. Trace out the degrees of freedom in the part B. Such a procedure can
be presented elegantly by a graphical notation described in Fig.~\ref%
{fig:MPS-Graph}(c).\ The contribution of the subsystem B is represented by
the transfer matrix $T=\sum_{p}B^{\left[ p\right] }\otimes \bar{B}^{\left[ p%
\right] }$. The expression $\rho _{A}$ can be written into a matrix product
operator form, which is displayed in Fig.~\ref{fig:MPS-Graph}(c)
\begin{eqnarray}
\rho _{A} &=&Tr\left( \prod_{j}R_{j}\right) , \\
R_{j} &=&\sum_{p_{j},q_{j}}|p_{j}\rangle \langle q_{j}|\left( B^{\left[ p_{j}%
\right] }\otimes \overline{B}^{\left[ q_{j}\right] }\right) T.
\end{eqnarray}

Step 3. To derive the EH, we have to express the projection operator $%
|p_{j}\rangle \langle q_{j}|$ in terms of product of spin operators. Note
that each $|p_{j}\rangle $ is composed of two spin-$S/2$'s and we can write
an expansion:$|m\rangle \langle n|=\sum_{i}\Gamma _{\left( m,n\right)
,i}O^{i}$, where\ $O^{i}$ ($i=0$,$1,...S^{2}+2S$) are the spin-$S/2$
operators with $O^{0}=I$. With these considerations, the full expression $%
R_{j}$ is written as{\small
\begin{eqnarray}
R_{j} &=&\sum_{\left\{ p_{j},q_{j}\right\} }\sum_{\left\{ m,n,\alpha
\right\} }\left[ \left( B^{\left[ p_{j}\right] }\otimes \overline{B}^{\left[
q_{j}\right] }\right) T\right] \chi _{m_{1},m_{2}}^{p_{j}}\overline{\chi }%
_{n_{1},n_{2}}^{q_{j}}  \notag \\
&&\times \Gamma _{\left( m_{1},n_{1}\right) ,\alpha _{2j-1}}\Gamma _{\left(
m_{2},n_{2}\right) ,\alpha _{2j}}O^{\alpha _{2j-1}}O^{\alpha _{2j}}.
\end{eqnarray}%
} It is emphasized that no approximation has been made so far.

Step 4. Since the form $R_{j}$ is complicated, a controlled approximation
can be introduced. For a long block length $l$, the only dominant coupling
in $\rho _{A}$ is $\delta =\left( \frac{-S}{S+2}\right) ^{l}$, which implies
that the exchange coupling between two edge spins decays exponentially. Then
$R_{j}$ can be separated into two individual edge spins, and the final
result for $H_{E}$ is given by
\begin{equation}
H_{E}\approx \frac{12}{S\left( S+2\right) }\left( \frac{-S}{S+2}\right)
^{l}\sum_{i}\mathbf{s}_{i}\cdot \mathbf{s}_{i+1},
\end{equation}%
where $\mathbf{s}_{i}$ is the fractionalized edge spin-$S/2$'s in the
reduced system.\ The detailed derivations for the $S=1$ and $S=2$ cases are
included in the supplementary material.

Therefore, the resulting entanglement properties can be divided into three
categories: (i) For $l=odd$, $H_{E}$ represents a ferromagnetic ordered
phase with spin wave excitations (the numerical result for $S=1$ is included
in the supplementary material); (ii) For $l=even$ and $S=even$, $H_{E}$
describes the Haldane gapped phase with integer spins. In particular, for $%
S=4n+2$ with integer $n$, it represents the SPT phase of the odd integer
spin Haldane phase even though the original VBS state corresponds to the
topologically \textit{trivial} state. (iii) For $l=even$ and $S=odd$, $H_{E}$
is just the quantum Heisenberg antiferromagnetic half-odd integer spin model
with quantum critical ground state\cite{Affleck-Haldane}. The corresponding
effective field theory for $S>1$ describes a multicritical point
characterized by the 1+1 (space-time) dimensional SU(2) level-$S$
Wess-Zumino-Witten (WZW) theory, but the stable fixed point of these
critical phases is determined by the SU(2) level-1 WZW theory\cite%
{Affleck-Haldane,Oshikawa}. These are the important properties encoded in
the AKLT VBS states with integer spins.

\section{Numerical calculations}

\subsection{Spin-1 AKLT state}

\begin{figure}[t]
\includegraphics[width=8cm]{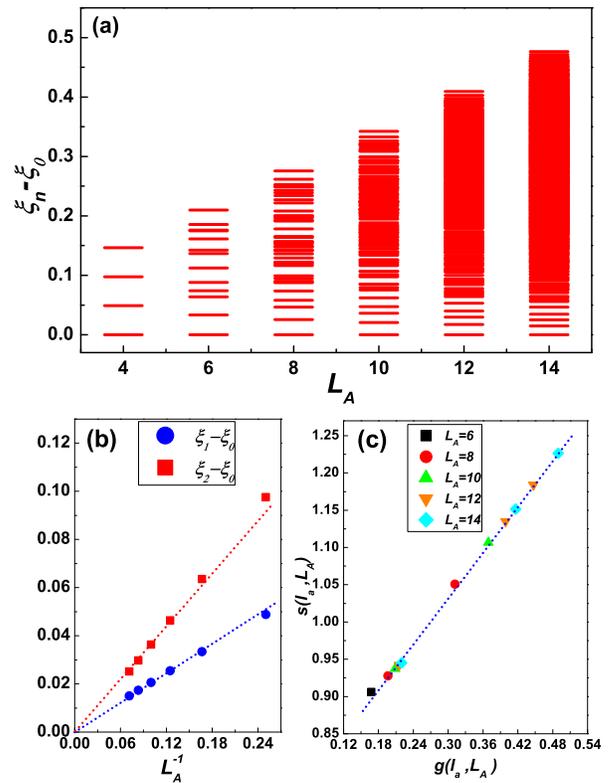}
\caption{(a) Bulk ES with $S=1$ and the block length $l=4$. (b) Two lowest
entanglement levels are linear with $L_{A}^{-1}$. (c) The entanglement
entropy $s(l_{a},l_{A})$ as a function of $g(l_{a},L_{A})=\frac{1}{3}\ln %
\left[ \frac{L_{A}}{\protect\pi }\sin \left( \frac{\protect\pi l_{a}}{L_{A}}%
\right) \right] $. }
\label{fig:Block4-BES}
\end{figure}

In order to put the above analytical results on a solid ground, we perform
the exact numerical diagonalization for the reduced density matrix $\rho
_{A} $ for the spin-$1$ AKLT state \textit{without} any approximations. The
full entanglement spectrum (ES) for the block length $l=4$ is displayed in
Fig.~\ref{fig:Block4-BES}(a). We use the effective length $L_{A}$ to denote
the reduced system length, independent of the block length in the original
scale. The degeneracies of each levels correspond to $1,3,5$,etc for every
system length. The density of entanglement entropy $S_{A}/L_{A}$ saturates
to $0.6929$ very quickly, close to the value $\ln 2=0.6931$. The
entanglement spectral gap\ $\xi _{1}-\xi _{0}$ is found to scale linearly
with the inverse subsystem length $\xi _{1}-\xi _{0}=k_{1}L_{A}^{-1}$ shown
in Fig.~\ref{fig:Block4-BES}(b), suggesting the bulk ES is gapless in the
thermodynamic limit. Moreover, the second excited entanglement level is also
fitted as $\xi _{2}-\xi _{0}=k_{2}L_{A}^{-1}$ displayed in Fig.~\ref%
{fig:Block4-BES}(b), and the ratio of these two excited levels is determined
as $k_{2}/k_{1}=1.975\sim 2$, implying the difference of scaling dimensions
for these two excited levels is $2$.

To determine the universality class of this spectrum, we focus on the wave
function of the lowest level $|\psi _{0}\rangle $. By further cutting the
reduced system into two halves with lengths $l_{a}$ and $\left(
L_{A}-l_{a}\right) $, respectively, we calculate the entanglement entropy: $%
s\left( l_{a},L_{A}\right) =$Tr$_{l_{a}+1,l_{a}+2..L_{A}}\left( |\psi
_{0}\rangle \langle \psi _{0}|\right) $. Fitting to the Calabrese-Cardy
formula\cite{Calabrese},
\begin{equation}
s\left( l_{a},L_{A}\right) =\frac{c}{3}\ln \left[ \frac{L_{A}}{\pi }\sin
\left( \frac{\pi l_{a}}{L_{A}}\right) \right] +s_{0},
\end{equation}%
we obtain the central charge $c=$ $1.02\pm 0.02$ in Fig.~\ref{fig:Block4-BES}%
(c). This result confirms that the obtained ES belongs to the universality
class of the 1+1 (space-time) dimensional $SU\left( 2\right) _{1}$ WZW
conformal field theory, which is the same as the quantum antiferromagnetic
Heisenberg spin-$1/2$ chain. The corresponding EH describes the critical
point separating the spin-1 Haldane phase from the trivial gapped phase\cite%
{Rao-Wan-Zhang}. Such a critical point differs the critical point between
the Haldane phase and dimerized phase in the SO(3) bilinear-biquadratic
spin-1 chain from that the dimerized phase has spontaneously translation
symmetry breaking\cite{Kitazawa}. It is a multicritical point described by
the 1+1 dimensional SU(2)$_{2}$ WZW theory with $c=3/2$.

\begin{figure}[t]
\includegraphics[width=8cm]{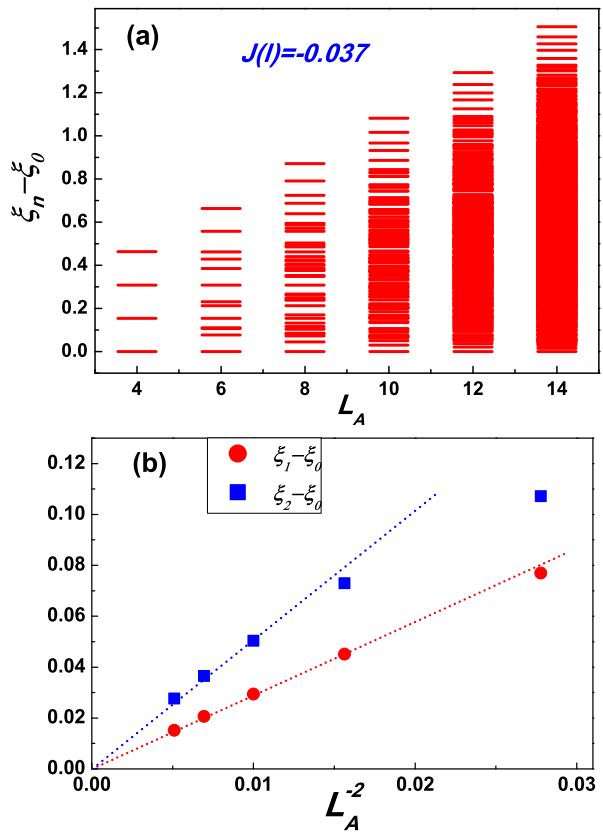}
\caption{(a) The bulk ES with $S=1$ and the block length $l=3$, the lowest 
level is $\left( L_{A}+1\right) $-fold degenerate. (b) The entanglement 
spectral gap is linear with square of the inverse subsystem size, 
indicating a spin-wave excitation for the ferromagnetic Heisenberg spin chain. 
The second excited level is also plotted, but the data from small sizes 
slightly deviate from the line.}
\label{fig:Block3}
\end{figure}

When the block includes the odd number of lattice sites, e.g. $l=3$, the
bulk ES is calculated and displayed in Fig.~\ref{fig:Block3}(a). The
entanglement entropy density is found to saturate to $0.691$, which is
within $0.3\%$ to the value of $\ln 2$. The lowest entanglement level $\xi
_{0}$ is linear with system size. However, the lowest entanglement level has
the degeneracy $L_{A}+1$ in each system size. We computed the magnetization
distribution $m_{tot}^{z}=\sum_{i}m_{i}^{z}$\ for these states and found
they are well located in $\left[ -L_{A}/2,L_{A}/2\right] $, indicating that
a large spin-$L_{A}/2$ is formed. This can only be achieved in the
ferromagnetic interacting between edge spins, and the bulk ES thus describes
a ferromagnetical long-range ordered state. For more evidence, we evaluate
the entanglement gap scales as $\xi _{1}-\xi _{0}\sim L_{A}^{-2}$, which is
a direct sign of spin-wave excitations. To further confirm the this
spectrum, we fit the second excitation level $\xi _{2}-\xi _{0}\sim
L_{A}^{-2}$, as dictated in Fig.~\ref{fig:Block3}(b). If we take the
Heisenberg interaction as the EH, the coupling constant is fitted to be $%
J\sim -0.037$, while in our analytical analysis it is $J=\left( -1/3\right)
^{3}=-0.038$. Thus, our analytical derivation is confirmed.

\subsection{Spin-2 AKLT state}

\begin{figure}[t]
\includegraphics[width=8cm]{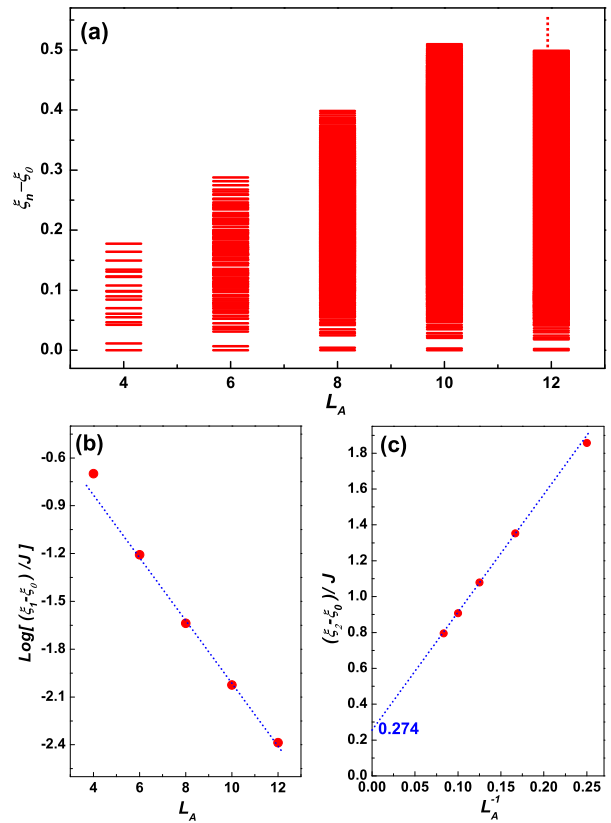} 
\caption{(a) The bulk ES with $S=2$ and the block length $l=6$ for an open
chain. The lowest level is singlet, and the first excited level is triplet.
(b) The first excited level decays exponentially with the subsystem size.
(c) The bulk excitation energy $\left( \protect\xi _{2}-\protect\xi %
_{0}\right) /J\left( l\right) $ saturates to a finite value in thermodynamic
limit.}
\label{fig:Spin2-obc}
\end{figure}

Another important calculation is performed for the spin-2 AKLT VBS state.
The ES with $l=6$ under \textit{open} boundary condition is presented in
Fig.~\ref{fig:Spin2-obc}(a). The lowest level is singlet and the first
excited level is triplet. However, the level spacing between these two
states is fitted as an exponential decay with the subsystem size: $\left(
\xi _{1}-\xi _{0}\right) /J\left( l\right) \sim e^{-L_{A}/\Delta }$ with $%
\Delta =4.769$, displayed in Fig.~\ref{fig:Spin2-obc}(b). Here the
antiferromagnetic Heisenberg coupling strength $J\left( l\right) $ is fitted
to be $0.0230$, very close to our analytical value $0.0234$. In the
thermodynamic limit, the lowest entanglement level becomes four-fold
degenerate. These results are consistent with the defining property of the
topological spin-1 Haldane phase. Moreover, $\left( \xi _{2}-\xi _{0}\right)
/J\left( l\right) $ shown in Fig.~\ref{fig:Spin2-obc}(c) approaches to the
finite value $0.274$, smaller than the Haldane gap value $0.41$ from the
density matrix renormalization group calculation\cite{white}. The difference
can be improved when the longer length of the effective spin chain is
calculated.

\section{Discussion and Conclusion}

The symmetric bulk bipartition allows us to establish a general description
of QCP separating the SPT phase from its trivial gapped phase \textit{%
directly} from the fixed point wave function of the topological phase. For
the one-dimensional SPT phase with the protecting symmetry of $G=SO(3)$ Lie
group, its fundamental group is $\Pi _{1}(G)=Z_{2}$. So there are only two
different phases: the odd integer spin Haldane gapped phase and its trivial
gapped phase adiabatically connected to the even integer spin Haldane gapped
phase. A QCP exists to separate these two phases, and the effective model
Hamiltonian for this QCP is just given by the quantum antiferromagnetic
Heisenberg half-odd integer spin chain. The corresponding critical theory is
characterized by the 1+1 (space-time) dimensional $SU(2)_{1}$ WZW conformal
field theory with the Lie group $\tilde{G}=SU(2)$, where $\tilde{G}$ is just
the universal covering group of $G$ and has a trivial fundamental group, $%
\Pi _{1}(\tilde{G})=1$. Our results may thus generalize the widely discussed
bulk-edge correspondence: the bulk topological property of the topological
phase not only determines its symmetry-protected edge degrees of freedom,
but also the critical properties of the second order phase transition to the
trivial phase. Furthermore, the fundamental degrees of freedom of the
critical theory are precisely these edge degrees of freedom \emph{confined}
in the bulk of the topological phase. As a result this QCP is a typical
deconfined critical point. These results can be generalized for other SPT
phases with the protecting symmetry of continuous Lie group.

To summarize, we have applied a symmetric bulk bipartition to the
one-dimensional AKLT VBS states for the integer spin-S Haldane gapped phase,
and an array of fractionalized spin-S/2 edge spins can be created in the
reduced bulk system. Via the calculations of the bulk entanglement spectra
for the reduced system, the topological properties encoded in the original
VBS wave functions are revealed.

\textit{Acknowledgements.- } {G. M. Zhang would like to thank D. H. Lee and
X. Wan for the helpful discussions and acknowledge the support of NSF-China
through Grant No.20121302227. K. Yang is supported by NSF grants DMP-1442366
and DMP-1157490.}

\end{document}